\RequirePackage[l2tabu, orthodox]{nag}

\documentclass[utf8,biblatex,english]{lni}
\bibliography{lni-paper-example-de.bib}

\usepackage{booktabs}

\usepackage[math]{blindtext}
\usepackage{mwe}
\usepackage{makecell}

\iffalse
\AtEveryBibitem{%
  \ifentrytype{article}{%
  }{%
    \clearfield{doi}%
    \clearfield{issn}%
    \clearfield{url}%
    \clearfield{urldate}%
  }%
  \ifentrytype{inproceedings}{%
  }{%
    \clearfield{doi}%
    \clearfield{issn}%
    \clearfield{url}%
    \clearfield{urldate}%
  }%
}
\fi

\begin{document}
\title[Automated Statement Extraction]{Automated Statement Extraction from Press Briefings}
\author[Jüri Keller \and Meik Bittkowski \and Philipp Schaer]
{Jüri Keller\footnote{Technische Hochschule Köln,
  Claudiusstraße 1, 50678 Köln, Germany \email{jueri.keller@smail.th-koeln.de}} \and
 Meik Bittkowski\footnote{Science Media Center Germany, Schloss-Wolfsbrunnenweg 33, 69118 Heidelberg, Germany \email{bittkowski@sciencemediacenter.de}}\and Philipp Schaer\footnote{Technische Hochschule Köln,
  Claudiusstraße 1, 50678 Köln, Germany \email{philipp.schaer@th-koeln.de}}}

\startpage{1} 
\editor{B. K{"o}nig-Ries et al.} 
\booktitle{Datenbanksysteme f{"u}r Business, Technologie und Web (BTW 2023)} 
\maketitle
\begin{abstract}
Scientific press briefings are a valuable information source. They consist of alternating expert speeches, questions from the audience and their answers. Therefore, they can contribute to scientific and fact-based media coverage. Even though press briefings are highly informative, extracting statements relevant to individual journalistic tasks is challenging and time-consuming.
To support this task, an automated statement extraction system is proposed. Claims are used as the main feature to identify statements in press briefing transcripts. The statement extraction task is formulated as a four-step procedure. First, the press briefings are split into sentences and passages, then claim sentences are identified through sequence classification. Subsequently, topics are detected, and the sentences are filtered to improve the coherence and assess the length of the statements. 

The results indicate that claim detection can be used to identify statements in press briefings. While many statements can be extracted automatically with this system, they are not always as coherent as needed to be understood without context and may need further review by knowledgeable persons.
\end{abstract}


\section{Introduction}

Scientific press briefings are a valuable instrument in scientific communication. They consist of alternating expert speeches and answers to questions from the audience, which let scientists and journalists immediately and jointly address the information needs of journalists. The SMC press briefings\footnote{\url{https://www.sciencemediacenter.de/alle-angebote/alle-angebote/?tx_solr\%5Bfilter\%5D\%5B0\%5D=type\%3APress+Briefing}} used in this work, typically begin with an introduction of the participating experts followed by moderated questions from journalists and answers from invited experts. Additionally, the key results are concluded in closing statements. Press briefings can directly contribute to a more fact-based media coverage by connecting journalists and scientists. Even though press briefings are highly informative, filtering this information and extracting relevant statements remains challenging and time-consuming due to the high entropy and domain-specific language used.

In this context, the research question will be answered: To what extent can claim detection be used to extract statements from scientific press briefings?

To approach this domain-specific problem and automatically identify relevant statements from press briefings, a pipeline is proposed, relying on claims as a central element of a statement.
A transformer based language model \cite{chanGermanNextLanguage2020} is used to classify claim sentences, which are subsequently filtered and clustered to improve the relevance and coherence of the emerging statements. 

For example, a sentence like \textit{"We do not yet know what a long covid course of Omicron infection looks like."} is considered a \textit{complete claim}, while the sentence \textit{"If I only have mild symptoms, it doesn't mean I won't have problems in the long run."} is only considered an \textit{incomplete claim} as it can not be understood without context.

The main contribution of this work is the application of state-of-the-art methods in claim detection to develop a system that extracts statements from transcripts of scientific press briefings. Additionally, a novel dataset of 53 transcribed German press briefings is created and partially annotated.
The dataset and implementation are made publicly available via GitHub\footnote{Press Briefing Claim Dataset at GitHub: \url{https://github.com/jueri/press_briefing_claim_dataset}}\footnote{Statement Extractor at GitHub: \url{https://github.com/jueri/statement_extractor}} and Zenodo \cite{keller_juri_2022_6047551}.

\section{Related Work}\label{sec:Related Work}
Automated extraction of statements from press briefings occurs at the intersection of journalism and computer science. While the premise and context emerge from journalism, the technical methods originate from computer science. 



Various definitions of a claim exist \cite{daxenbergerWhatEssenceClaim2017, liebeckWhatAirportMining2016}. Generically, a claim indicates the author's position related to a main concept (topic) of a sentence \cite{levyUnsupervisedCorpuswideClaim2017}. Therefore, claims are indicated as the main feature of a statement. Generically, a claim is a short phrase people can have different opinions on \cite{liebeckWhatAirportMining2016, lippiArgumentationMiningState2016}. Since claims are conceptually complex, it is often difficult to determine what a claim is and what not \cite{lippiContextindependentClaimDetection2015}. To clarify, this work refers to a claim as a sentence asserting something on the main concept.

Claim detection is an information extraction task often part of a data processing pipeline solving more complex problems. Automatically detecting claims in texts has a variety of applications in fields like decision-making, argument mining, fact-checking, or document processing \cite{levyArgumentativeContentSearch2018, daxenbergerWhatEssenceClaim2017}. As diverse as the applications for claim detection are, so are the domains it is applied in, such as for example political, debate, legal, and the web \cite{lippiArgumentationMiningState2016}.

Besides binary classification, different claim types or the part of the sentence that is a claim can be identified \cite{konstantinovskiyAutomatedFactcheckingDeveloping2018, lippiArgumentationMiningState2016}. Levy et al. \cite{levyContextDependentClaim2014} first formally defined the task of automated claim detection for computational argumentation while working on automatic argument mining \cite{levyContextDependentClaim2014, slonimAutonomousDebatingSystem2021}. The systems developed in the following years can be categorized by their ability to detect claims without further context or the scope of application, e.g., a corpus of documents or a single one \cite{lippiContextindependentClaimDetection2015, daxenbergerWhatEssenceClaim2017}. Chakrabarty et al. and Daxenberger et al. provide more extensive overviews of different claim detection models and datasets  \cite{chakrabartyIMHOFinetuningImproves2019, daxenbergerWhatEssenceClaim2017}.

\section{Statement Detection System}\label{sec:Statement Detection System}
\begin{figure}[t]
\centering
	\includegraphics[scale=0.4]{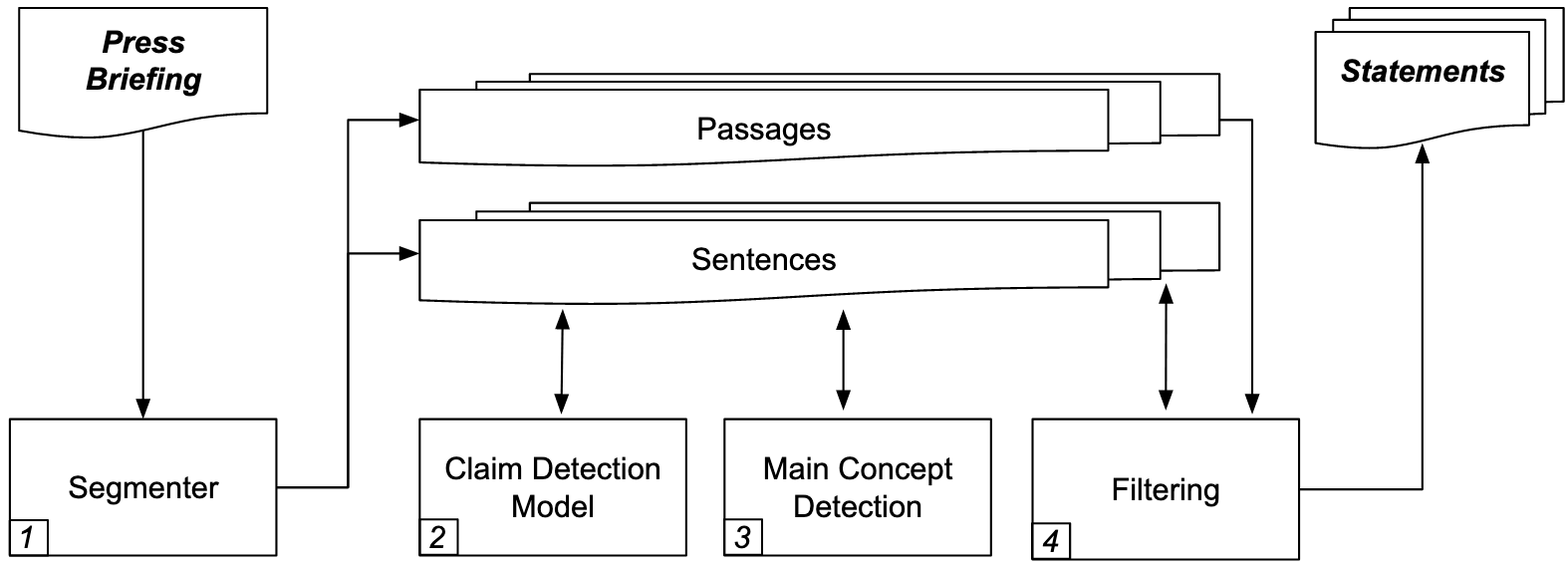}
 	\begin{footnotesize}
 		\caption{Schematically visualization of the pipeline approach.}
 		\label{fig:map}
 	\end{footnotesize}
 \end{figure}

Figure \ref{fig:map} visualizes the pipeline approach to extract statements from press briefings. The press briefing needs to be initially segmented into passages (1). Then, at the system's core, a classification model is used to identify sentences containing claims (2). Further, the main concept of the press briefing and the sentences are detected (3). This information is used in the last step to filter out the sentences that are not or incomplete claims (4).

\textbf{1. Segmentation:} The press briefings are split by a sentence tokenizer, and neighboring and coherent sentences are combined into statements of multiple sentences. Therefore, the sentences are represented in vector space through spaCy \cite{SPACY} and neighboring sentences are clustered by similarity with the TextSplit \cite{alemiTextSegmentationBased2015} algorithm. Starting with a list of vector representations of sentences, TextSplit calculates a score for each segment. By modifying the segment boundaries, the score changes. The optimal segmentation is determined by minimizing the aggregated segment scores \cite{alemiTextSegmentationBased2015, modyFindingBestPart2020}. 

\textbf{2. Claim Detection:} A German language model \cite{chanGermanNextLanguage2020} based on the BERT \cite{devlinBERTPretrainingDeep2019} architecture is fine-tuned for the task of sentence classification. The model's hyperparameters are tuned to achieve the best results on available training data. The German BERT version GBERT is used as the foundational model for the classifier \cite{chanGermanNextLanguage2020}. This model is fine-tuned as a claim sentence classifier using a dataset of 3000 sentences created for this purpose from the publicly available SMC press briefings\footnote{\url{https://www.sciencemediacenter.de/alle-angebote}}. The dataset contains a total of 53 press briefings from a time span of over four years. It consists of 24,897 sentences with an average length of 17.31 tokens. 


\label{sec:Main Concept Detection}
\textbf{3. Main Concept Detection:} Based on the intuitive assertion that a claim sentence with a topic similar to the overall topic of the press briefing the sentence originates from is most relevant, the topics of both the sentences and the overall press briefings (i.e., the title and the introduction text) are detected.
Therefore, Wikipedia articles are used as representations for the topic of a sentence or the whole press briefing. All Wikipedia articles related to tokens in the sentences and the title and the introduction texts are detected. The sentence is considered more relevant if a sentence can be linked to the same Wikipedia articles as the title or introduction text. For this task, the wikification APIs Dandelion\footnote{\url{https://dandelion.eu/}} and TagMe\footnote{\url{https://sobigdata.d4science.org/web/tagme/tagme-help}} are used \cite{ferraginaTAGMEOntheflyAnnotation2010}.

\textbf{4. Sentence Filtering:} Two techniques are applied to filter out claim sentences with a low topical similarity between the main concept of the sentence and the overall main concept to improve the coherence and relevance. The similarity is measured by two approaches based on embeddings or Wikipedia concepts. Both approaches create a similarity score that can be used for filtering with a minimum threshold.

The first approach calculates similarity by creating a vector representation from the title of the press briefing and the individual sentences and then measuring the cosine similarity between both. The vector representations are created by combining spaCy \cite{SPACY} word embeddings.
The second approach is based on the Wikipedia concepts detected as described previously. The confidence scores of shared Wikipedia concepts between the title or introduction text and the individual sentence are summed to create a score of topical relatedness.

\section{Experimental Evaluation}\label{sec:Results}
Three experiments were conducted to evaluate the system's ability to extract statements from press briefings. Therefore, three new press briefings with 799 sentences from different categories were annotated as ground truth.

\textbf{Claim Detection:} 
For hyperparameter optimization, models with different configurations were trained for six epochs. The optimal number of epochs was determined in conjunction with the learning rate by comparing the evaluation loss rates after each epoch. Learning rates, ranging from 0.00005 to 0.01, were tested. The best results could be achieved with a learning rate of 1e-5 and three epochs. For the final system, a model with an F1 Score of 0.89, a precision of 0.92 and a recall of 0.86 was used. This model assigned slightly more false positive than false negative labels, but since the human in the loop can decide on the misclassified sentences, this behavior is preferred.

To evaluate the claim detection component of the system, claim sentences classified with minimal confidence scores of 0.7, 0.8 and 0.9 are analyzed independently. Since the confidence score limits the number of sentences considered as claims by the model, a score small enough to allow claim sentences to be classified and high enough to exclude non-claim sentences leads to the best results.
The overall best results for detecting claims independent of their completeness were achieved at a confidence of 0.7 with an F1 score of 0.68, a precision of 0.89 and a recall of 0.55.
By investigating individual results for the claim types, \textit{complete claims} (that can stand on their own) and \textit{incomplete claims} (that can not), it can be assessed if the filtering and sentence clustering methods improve the coherence. In total, a maximum of 167 of 224 \textit{incomplete claims} and 102 of 152 \textit{complete claims} claim sentences could be identified with various system configurations. With the claim detection model and minimum confidence of 0.8, the highest F1 score of 0.47 could be achieved. Raising the confidence threshold to 0.9 results in a higher precision of 0.43. Decreasing the score to 0.7 leads to a higher recall of 0.67.

\textbf{Statement Filtering:} Based on the results of the first experiments, the claim sentences with a confidence score of 0.8 are further filtered as this threshold achieved the best F1 score. To exclude incoherent statements, the similarity between the sentence's main concept and the press briefing's main concept is calculated. The main concepts of the press briefing based on \textit{embeddings}, \textit{title wikification} and \textit{introduction wikification} are evaluated individually. For each method, a minimum similarity is chosen by investigating the similarity distribution for bends.
With the \textit{introduction wikification} based filtering, a precision of 0.43 was achieved. The \textit{title wikification} method reaches a precision of 0.456, and the \textit{embedding} method exceeds this score with the best precision of 0.463. All results are presented in table \ref{tab:complete_claims}.

\begin{table}[t]
    \centering
        \caption{Results for detected \textit{complete claims}. The first three columns hold the results for different confidence thresholds of the claim detection model. The last three columns contain the results of a system using a claim detection model threshold of 0.8 and the different main concept methods. The best results are highlighted in \textbf{bold}.}
  \label{tab:complete_claims}
  \begin{tabular}{rrrrrrr}
    \toprule
{Confidence} & 0.9 & 0.8 & 0.7 & \makecell{0.8 \\embedding} & \makecell{0.8 \\w. title} & \makecell{0.8 \\w. intro}\\
    \midrule
F1 & 0.466 & \textbf{0.473} & 0.450 & \textbf{0.481} & 0.339 & 0.341 \\ 
Precision & \textbf{0.426} & 0.378 & 0.339 & \textbf{0.463} & 0.456 & 0.430 \\ 
Recall & 0.513 & 0.632 & \textbf{0.671} & \textbf{0.500} & 0.270 & 0.283 \\ 
    \bottomrule
  \end{tabular}
\end{table}

\textbf{Sentence Clustering:} The last experiment evaluates the sentence clustering method. Claim sentences with a confidence of at least 0.9 are enlarged to statements of multiple sentences. The statements are then assessed according to their coherence. To measure if the larger statements created by the similarity-based sentence clustering are more coherent, the \textit{complete claim} and \textit{incomplete claim} ratio for all methods are compared. With 53 \% \textit{complete claims}, the coherence of the statements extracted exceeds all other methods and the baseline. 46 \% of the statements extracted with the claim detection module and a minimum confidence score of 0.9 are complete statements. The embedding method of the main concept module extracts 50 \% \textit{complete claims}.

\section{Discussion and Conclusion}\label{sec:Discussion_Conclusion}
The analysis shows that the press briefings are very argumentative content full of claims. While most claims are incomplete and can only be understood with further information, 46 \% can stand alone. The sentences extracted by the proposed system are highly likely to be claims, although not necessarily complete claims. This finding is supported by the low false-positive rate of the system considering all claims. Missing coherence is the main error of the statements extracted.

Compared to the initial results of the claim detection model with an accuracy of 0.89, the results achieved in the experiments are noticeably lower. This may be caused by the dataset used for training, which only contains \textit{complete claims} and \textit{no claims} and should be improved in future work. 
By adding a filter based on main concepts, some incoherent claim sentences could be successfully excluded, and the precision improved. A smaller confidence threshold of the initial claim detection module could improve these results further by providing more claim sentences in general. The embedding-based main concept method provided the best results. The Wikification-based methods come to an extent as some topics can only be expressed poorly by Wikipedia concepts, and other Wikipedia concepts may not exist or be captured by the model.

Assessing statements of multiple sentences, some statements previously classified as \textit{incomplete claims} can be understood without additional information through the added neighboring sentences. By systematically adding sentences, more coherent statements could be created. However, this method sacrifices the conciseness of statements since they get longer.

In summary, this work investigates how statements can be mined from scientific press briefings. A dataset is created from publicly available press briefings and used to fine-tune a claim detection model. Additional methods are tested to improve the quality of the extracted statements. More context is added to the statements, and the statements are filtered based on their topic. The results are compared to a gold standard to evaluate the system's performance.

The results show that the system can differentiate between claims of any type and \textit{no claim} sentences with high precision (0.89), but only half of the claims can be recalled (0.55). Considering only \textit{complete claim} sentences, the system's overall performance decreases (F1 0.47), but relatively more complete claims can be identified. These results emphasize the system's difficulty differentiating between complete and incomplete claim sentences. The resulting statements gain coherence by adding more context. Similarly, incomplete claim sentences can be filtered out by low topical similarity, which improves the results slightly.
The evaluations show that the presence of a claim can be used as an indicator to detect statements. However, the results leave room for improvement in many aspects. Especially the coherence of the extracted statements needs to be improved to extract statements that can stand on their own.


\printbibliography

@article{slonimAutonomousDebatingSystem2021,
	title = {An autonomous debating system},
	volume = {591},
	issn = {0028-0836, 1476-4687},
	url = {http://www.nature.com/articles/s41586-021-03215-w},
	doi = {10.1038/s41586-021-03215-w},
	language = {en},
	number = {7850},
	urldate = {2021-10-20},
	journal = {Nature},
	author = {Slonim, Noam and Bilu, Yonatan and Alzate, Carlos and Bar-Haim, Roy and Bogin, Ben and Bonin, Francesca and Choshen, Leshem and Cohen-Karlik, Edo and Dankin, Lena and Edelstein, Lilach and Ein-Dor, Liat and Friedman-Melamed, Roni and Gavron, Assaf and Gera, Ariel and Gleize, Martin and Gretz, Shai and Gutfreund, Dan and Halfon, Alon and Hershcovich, Daniel and Hoory, Ron and Hou, Yufang and Hummel, Shay and Jacovi, Michal and Jochim, Charles and Kantor, Yoav and Katz, Yoav and Konopnicki, David and Kons, Zvi and Kotlerman, Lili and Krieger, Dalia and Lahav, Dan and Lavee, Tamar and Levy, Ran and Liberman, Naftali and Mass, Yosi and Menczel, Amir and Mirkin, Shachar and Moshkowich, Guy and Ofek-Koifman, Shila and Orbach, Matan and Rabinovich, Ella and Rinott, Ruty and Shechtman, Slava and Sheinwald, Dafna and Shnarch, Eyal and Shnayderman, Ilya and Soffer, Aya and Spector, Artem and Sznajder, Benjamin and Toledo, Assaf and Toledo-Ronen, Orith and Venezian, Elad and Aharonov, Ranit},
	month = mar,
	year = {2021},
	pages = {379--384},
}

@inproceedings{ferraginaTAGMEOntheflyAnnotation2010,
	address = {Toronto, ON, Canada},
	title = {{TAGME}: on-the-fly annotation of short text fragments (by wikipedia entities)},
	isbn = {978-1-4503-0099-5},
	shorttitle = {{TAGME}},
	url = {http://portal.acm.org/citation.cfm?doid=1871437.1871689},
	doi = {10.1145/1871437.1871689},
	language = {en},
	urldate = {2021-10-07},
	booktitle = {Proceedings of the 19th {ACM} international conference on {Information} and knowledge management - {CIKM} '10},
	publisher = {ACM Press},
	author = {Ferragina, Paolo and Scaiella, Ugo},
	year = {2010},
	pages = {1625},
}

@misc{modyFindingBestPart2020,
	title = {Finding the best part of your podcast to promote via {NLP}},
	url = {https://towardsdatascience.com/finding-the-best-part-of-your-podcast-to-promote-via-nlp-f844a88b287a},
	language = {en},
	urldate = {2021-09-08},
	journal = {Medium},
	author = {Mody, Neil},
	month = jun,
	year = {2020},
}

@inproceedings{chakrabartyIMHOFinetuningImproves2019,
	title = {{IMHO} fine-tuning improves claim detection},
	url = {https://doi.org/10.18653/v1/n19-1054},
	doi = {10.18653/v1/n19-1054},
	booktitle = {Proceedings of the 2019 conference of the north american chapter of the association for computational linguistics: {Human} language technologies, {NAACL}-{HLT} 2019, minneapolis, {MN}, {USA}, june 2-7, 2019, volume 1 (long and short papers)},
	publisher = {Association for Computational Linguistics},
	author = {Chakrabarty, Tuhin and Hidey, Christopher and McKeown, Kathy},
	editor = {Burstein, Jill and Doran, Christy and Solorio, Thamar},
	year = {2019},
	pages = {558--563},
}

@inproceedings{levyArgumentativeContentSearch2018,
	title = {Towards an argumentative content search engine using weak supervision},
	url = {https://aclanthology.org/C18-1176/},
	booktitle = {Proceedings of the 27th international conference on computational linguistics, {COLING} 2018, santa fe, new mexico, {USA}, august 20-26, 2018},
	publisher = {Association for Computational Linguistics},
	author = {Levy, Ran and Bogin, Ben and Gretz, Shai and Aharonov, Ranit and Slonim, Noam},
	editor = {Bender, Emily M. and Derczynski, Leon and Isabelle, Pierre},
	year = {2018},
	pages = {2066--2081},
}

@inproceedings{devlinBERTPretrainingDeep2019,
	title = {{BERT}: {Pre}-training of deep bidirectional transformers for language understanding},
	url = {https://doi.org/10.18653/v1/n19-1423},
	doi = {10.18653/v1/n19-1423},
	booktitle = {Proceedings of the 2019 conference of the north american chapter of the association for computational linguistics: {Human} language technologies, {NAACL}-{HLT} 2019, minneapolis, {MN}, {USA}, june 2-7, 2019, volume 1 (long and short papers)},
	publisher = {Association for Computational Linguistics},
	author = {Devlin, Jacob and Chang, Ming-Wei and Lee, Kenton and Toutanova, Kristina},
	editor = {Burstein, Jill and Doran, Christy and Solorio, Thamar},
	year = {2019},
	pages = {4171--4186},
}

@inproceedings{levyUnsupervisedCorpuswideClaim2017,
	title = {Unsupervised corpus-wide claim detection},
	url = {https://doi.org/10.18653/v1/w17-5110},
	doi = {10.18653/v1/w17-5110},
	booktitle = {Proceedings of the 4th workshop on argument mining, {ArgMining}@{EMNLP} 2017, copenhagen, denmark, september 8, 2017},
	publisher = {Association for Computational Linguistics},
	author = {Levy, Ran and Gretz, Shai and Sznajder, Benjamin and Hummel, Shay and Aharonov, Ranit and Slonim, Noam},
	editor = {Habernal, Ivan and Gurevych, Iryna and Ashley, Kevin D. and Cardie, Claire and Green, Nancy and Litman, Diane J. and Petasis, Georgios and Reed, Chris and Slonim, Noam and Walker, Vern R.},
	year = {2017},
	pages = {79--84},
}

@inproceedings{daxenbergerWhatEssenceClaim2017,
	title = {What is the essence of a claim? {Cross}-domain claim identification},
	url = {https://doi.org/10.18653/v1/d17-1218},
	doi = {10.18653/v1/d17-1218},
	booktitle = {Proceedings of the 2017 conference on empirical methods in natural language processing, {EMNLP} 2017, copenhagen, denmark, september 9-11, 2017},
	publisher = {Association for Computational Linguistics},
	author = {Daxenberger, Johannes and Eger, Steffen and Habernal, Ivan and Stab, Christian and Gurevych, Iryna},
	editor = {Palmer, Martha and Hwa, Rebecca and Riedel, Sebastian},
	year = {2017},
	pages = {2055--2066},
}

@article{lippiArgumentationMiningState2016,
	title = {Argumentation mining: {State} of the art and emerging trends},
	volume = {16},
	url = {https://doi.org/10.1145/2850417},
	doi = {10.1145/2850417},
	number = {2},
	journal = {ACM Trans. Internet Techn.},
	author = {Lippi, Marco and Torroni, Paolo},
	year = {2016},
	pages = {10:1--10:25},
}

@inproceedings{liebeckWhatAirportMining2016,
	title = {What to do with an airport? {Mining} arguments in the german online participation project tempelhofer feld},
	url = {https://doi.org/10.18653/v1/w16-2817},
	doi = {10.18653/v1/w16-2817},
	booktitle = {Proceedings of the third workshop on argument mining, hosted by the 54th annual meeting of the association for computational linguistics, {ArgMining}@{ACL} 2016, august 12, berlin, germany},
	publisher = {The Association for Computer Linguistics},
	author = {Liebeck, Matthias and Esau, Katharina and Conrad, Stefan},
	year = {2016},
}

@inproceedings{lippiContextindependentClaimDetection2015,
	title = {Context-independent claim detection for argument mining},
	url = {http://ijcai.org/Abstract/15/033},
	booktitle = {Proceedings of the twenty-fourth international joint conference on artificial intelligence, {IJCAI} 2015, buenos aires, argentina, july 25-31, 2015},
	publisher = {AAAI Press},
	author = {Lippi, Marco and Torroni, Paolo},
	editor = {Yang, Qiang and Wooldridge, Michael J.},
	year = {2015},
	pages = {185--191},
}

@article{alemiTextSegmentationBased2015,
	title = {Text segmentation based on semantic word embeddings},
	volume = {abs/1503.05543},
	url = {http://arxiv.org/abs/1503.05543},
	journal = {CoRR},
	author = {Alemi, Alexander A. and Ginsparg, Paul},
	year = {2015},
}

@inproceedings{levyContextDependentClaim2014,
	title = {Context dependent claim detection},
	url = {https://aclanthology.org/C14-1141/},
	booktitle = {{COLING} 2014, 25th international conference on computational linguistics, proceedings of the conference: {Technical} papers, august 23-29, 2014, dublin, ireland},
	publisher = {ACL},
	author = {Levy, Ran and Bilu, Yonatan and Hershcovich, Daniel and Aharoni, Ehud and Slonim, Noam},
	editor = {Hajic, Jan and Tsujii, Junichi},
	year = {2014},
	pages = {1489--1500},
}

@article{konstantinovskiyAutomatedFactcheckingDeveloping2018,
	title = {Towards automated factchecking: {Developing} an annotation schema and benchmark for consistent automated claim detection},
	volume = {abs/1809.08193},
	url = {http://arxiv.org/abs/1809.08193},
	journal = {CoRR},
	author = {Konstantinovskiy, Lev and Price, Oliver and Babakar, Mevan and Zubiaga, Arkaitz},
	year = {2018},
}

@inproceedings{chanGermanNextLanguage2020,
	title = {German's next language model},
	url = {https://doi.org/10.18653/v1/2020.coling-main.598},
	doi = {10.18653/v1/2020.coling-main.598},
	booktitle = {Proceedings of the 28th international conference on computational linguistics, {COLING} 2020, barcelona, spain (online), december 8-13, 2020},
	publisher = {International Committee on Computational Linguistics},
	author = {Chan, Branden and Schweter, Stefan and Möller, Timo},
	editor = {Scott, Donia and Bel, Núria and Zong, Chengqing},
	year = {2020},
	pages = {6788--6796},
}

@software{keller_juri_2022_6047551,
  author       = {Keller, Jüri},
  title        = {Automated Statement Extraction from Press Briefings},
  month        = feb,
  year         = 2022,
  publisher    = {Zenodo},
  version      = {0.1},
  doi          = {10.5281/zenodo.6047551},
  url          = {https://doi.org/10.5281/zenodo.6047551}
}

@software{SPACY,
  author       = {Matthew Honnibal and
                Ines Montani and
                Sofie Van Landeghem
                Adriane Boyd
                },
  title        = {spaCy: Industrial-strength Natural Language Processing in Python},
  month        = feb,
  year         = 2020,
  publisher    = {Zenodo},
  version      = {0.1},
  doi          = {10.5281/zenodo.1212303},
  url          = {https://doi.org/10.5281/zenodo.1212303}
}

\end{document}